\magnification=1100

\hsize 17truecm
\vsize 23truecm

\font\twelvec=msbm10 at 12pt
\font\sevenc=msbm10 at 9pt
\font\fivec=msbm10 at 7pt

\newfam\co
\textfont\co=\twelvec
\scriptfont\co=\sevenc
\scriptscriptfont\co=\fivec

\def\deg{\mathop{\rm deg}\nolimits}
\def\det{\mathop{\rm det}\nolimits}
\def\exp{\mathop{\rm exp}\nolimits}

\def\Id{\mathop{\rm Id}\nolimits}

\def\ker{\mathop{\rm Ker}\nolimits}
\def\lim{\mathop{\rm lim}\nolimits}

\def\Tra{\mathop{\rm Tr}\nolimits}
\def\tra{\mathop{\rm tr}\nolimits}

\def\ssp{\mathop{\rm sp}\nolimits}
\def\su{\mathop{\rm su}\nolimits}
\def\Sum{\displaystyle\sum}

\baselineskip 15pt

\centerline{\bf QUANTUM VORTICITY AT POSITIVE TEMPERATURE FOR SPINS SYSTEMS}

\centerline{\bf WITH CONTINUOUS SYMMETRY}
\medskip
\centerline{D.MINENKOV ${}^{(1)}$ and M.ROULEUX ${}^{(2)}$}
\medskip
\centerline {${}^{(1)}$ Institute for Problems in Mechanics of Russian Academy of Sciences}

\centerline {Prosp. Vernadskogo 101-1, Moscow, 119526, Russia, minenkov.ds@gmail.com}
\medskip
\centerline{${}^{(2)}$ Aix Marseille Univ, Univ Toulon, CNRS, CPT, Marseille, France, rouleux@univ-tln.fr} 
\bigskip
\noindent {\bf Abstract}: We propose a definition of vorticity at inverse temperature $\beta$ for Gibbs states in quantum XY or Heisenberg spin systems
on the lattice by testing $\exp[-\beta H]$ on
a complete set of observables (``one-point functions''). Imposing a compression of Pauli matrices
at the boudary, which stands for the classical environment, we perform some numerical simulations on finite lattices in case of XY model,
which exhibit usual vortex patterns.
\bigskip
\noindent {\bf 0. Introduction}.
\medskip
Consider the quantum XY or Heisenberg spin model for $S=1/2$ on the 2-D lattice ${\bf Z}^2$, with nearest neighbor interactions.
Marmin-Wagner, and Hohenberg theorems tell us that Gibbs states, for all inverse temperature $\beta$, are invariant
under simultaneous rotation of spins (absence of continuous symmetry breaking in two dimensions).
In the classical case, we know a bit more~: although there is a unique Gibbs state, with rotational symmetry,
which rules out the existence of first order transitions, a particular form for phase transition exists,
characterized by a change of behavior in the correlation functions. For the 2-D rotator, it has been
described by Berezinskii, and Kosterlitz-Thouless in term of topological excitations, called vortices [FrSp].
For the Heisenberg model, we observe higher order topological defects, called instantons  [BePo].

In this report, we make a first attempt to answer the natural question~: {\it How can we observe vorticity in the quantum case?}

Let us first consider a system in finite volume $\Lambda\subset{\bf Z}^2$.
The Hamiltonians we are interested in are of the form $H_\Lambda(\Phi)=-\Sum_{X\subset\Lambda}\Phi(X)$, where
$\Phi$ is an ``interaction'' between sites in $\Lambda$.
For nearest neighbor interaction, the contributing $X$ are pairs $\langle i,j\rangle$, and the Hamiltonian reads
$$H_\Lambda=-{1\over2}\Sum_{\langle i,j\rangle\subset\Lambda}(\sigma_i^x\otimes\sigma_j^x+\sigma_i^y\otimes\sigma_j^y+u\sigma_i^z\otimes\sigma_j^z)\leqno(0.1)$$
where $u$ is a coupling constant,
$u>0$ in the ferromagnetic case ($u=1$ when isotropic), $u<0$ in the anti-ferromagnetic case ($u=-1$ when isotropic), and $u=0$ is the XY model.
We could also add an external magnetic field $h\Sum_{i\in\Lambda}\sigma^z_i$ to $H_\Lambda$, 
and in case $u=0$, allow for anisotropy between $x$ and $y$ components. 

Though vortices can merge spontenaously in infinite volume, there are external fields that would certainly enhance vorticity.
General external fields are defined within the notion of a ``state'' [Si,II.3]. 

Throughout we denote by Tr the ordinary trace, and by tr the normalized trace, as $\tra (A)={1\over d}\Tra (A)$, where $d$ is the dimension.
Recall from [Si,II,1] the partial trace: if $A$
is a linear operator on ${\cal K}_1\otimes{\cal K}_2$,
the partial trace $\tra _{{\cal K}_1}$ or simply $\tra _1$ is an operator 
${\cal L}({\cal K}_1\otimes{\cal K}_2)\to{\cal L}({\cal K}_1)$ defined by the
requirement~:
$$\tra _{{\cal K}_1}\bigl(B({\tra}_1(A))\bigr)= {\tra}_{{\cal K}_1\otimes{\cal K}_2}((B\otimes1)A), \quad B\in{\cal L}({\cal K}_1)\leqno(0.2)$$

A quantum state $\rho$ assigns to each finite $X\subset{\bf Z}^2$ an operator $\rho_X$ with $\Tra (\rho_X)=1$ and 
$\tra _{{\cal H}_Y}(\rho_{X\cup Y})=\rho_X$
on ${\cal H}_Y=\otimes_{i\in Y}{\bf C}^2_i$ for all disjoint $X,Y\subset{\bf Z}^2$. 
(Instead of $\tra _{{\cal H}_Y}(A)$, we use also the notation $\tra _Y(A)$.~)
Given a state and the finite
interaction $\Phi(X)$, we define the Hamiltonian on all of ${\bf Z}^2$
$$H^\rho_\Lambda(\Phi)=-\Sum_{X\cap\Lambda\neq\emptyset}\Tra _{X\setminus\Lambda}
\bigl[(1\otimes\rho_{X\setminus\Lambda})\Phi(X)\bigr]\leqno(0.3)$$
that couples $\Lambda$ with the external field $\rho$ through its nearest neighbors at the boundary. 
For $A$ a quasi-local observable on ${\bf Z}^2$, we define the expectation value
$$\langle A\rangle^\rho_{\beta,\Lambda}={\Tra _\Lambda\bigl(\exp\bigl[-\beta H^\rho_\Lambda(\Phi)\bigr]A\bigr)\over 
\Tra _\Lambda\exp\bigl[-\beta H^\rho_\Lambda(\Phi)\bigr]}\leqno(0.4)$$
and $\langle A\rangle^\rho_{\beta,\Lambda}$ has a limit as $|\Lambda|\to\infty$. Such a state has been constructed 
in [AsPi] for the XY chain.

From a practical point of vue however, it is suitable to produce explicit ``approximate states'' that will 
favour the existence of
vortices in finite volume; we proceed in the following way.
Let $\Lambda\subset{\bf Z^2}$ be the ``small system'', 
and $\partial\Lambda\subset{\bf Z^2}$ its ``environment'', both finite. On $\partial \Lambda$, we ``compress'' the spin operators,
so that the measure of observable ``direction of spin'' is deterministic for $j\in\partial\Lambda$, 
and quantum for $j\in\Lambda$. The resulting Hamiltonian $H_{\Lambda\cup\partial\Lambda}(\Phi)$ accounts for 
the interaction with the approximate ``external field''
on $\partial\Lambda$, as in (0.3). 

Now, in finite volume $\Lambda\cup\partial\Lambda$, the only (normalized) Gibbs state is given by
$$A\mapsto\omega_\beta(A)={\tra (e^{-\beta H_{\Lambda\cup\partial\Lambda}} A)\over \tra (e^{-\beta H_{\Lambda\cup\partial\Lambda}}) }\leqno(0.5)$$
and called  the ``canonical Gibbs state''.  We shall actually define vorticity at inverse temperature $\beta$
by decomposing the linear form $\omega_\beta$ on a canonical (orthonormal) basis of observables.
\medskip
\noindent {\bf 1. Vorticity matrices}
\smallskip
Gibbs state (0.1) for spin $S=1/2$ systems, as a linear form on the ${\bf C}^*$-algebra of observables
$${\cal O}=\otimes_{j\in\Lambda\cup\partial\Lambda}o_j, \quad  o_j={\cal M}_{2\times2}({\bf C})$$
(``quasi-local observables'' if we were to consider the
thermodynamical limit,~) can be decomposed in a canonical basis. The simplest way is to restrict
to ``one-point functions'', i.e. the set $\widetilde{\cal O}\subset{\cal O}$
of $2N\times2N$, block-diagonal $2\times2$ matrices (Pauli matrices are spin representations of SU(2) of dimension 
$2S+1=2$), supported on individual sites of
$\Lambda\cup\partial\Lambda$,  $N=|\Lambda\cup\partial\Lambda|$.
\medskip
\noindent {\it a) The XY model}
\smallskip
We first consider the XY model, where we can restrict to real $2\times2$ matrices. 
The compression of Pauli matrices on $\partial\Lambda$ is given by the orthogonal projector 
$$\Pi(\theta)=\pmatrix{\cos^2{\theta}&\sin{\theta}\cos{\theta}\cr
\sin{\theta}\cos{\theta}&\sin^2{\theta}\cr}\leqno(1.1)$$ 
where $\theta$ parametrizes a point on the unit circle, accounting for the prescribed direction of ``vorticity'' on $\partial\Lambda$.
Let again
$\widetilde{\cal O}_{\bf R}\subset\widetilde{\cal O}$ be a real sub-algebra
$\widetilde{\cal O}$, of real dimension $4N$.
\medskip
\noindent {\it Example 1}:
$\widetilde{\cal O}_{\bf R}$ is the ``canonical'' algebra, generated by real matrices $(D^i)_{i\in\Lambda\cup\partial\Lambda}$,
whose all non-diagonal $2\times2$ blocks vanish, and all diagonal $2\times2$ blocks vanish, except
this supported on site $i$ that takes values in
$\{\delta_1,\delta_2,\delta_3,\delta_4\}$, where
$$\delta_1=\pmatrix{1\kern 1pt &0\kern 1pt \cr
0\kern 1pt & 0\kern 1pt }, \quad \delta_2=\pmatrix{0\kern 1pt &1\kern 1pt \cr
0\kern 1pt & 0\kern 1pt }, \quad \delta_3=\pmatrix{0\kern 1pt &0\kern 1pt \cr
1\kern 1pt & 0\kern 1pt }, \quad \delta_4=\pmatrix{0\kern 1pt &0\kern 1pt \cr
0\kern 1pt & 1\kern 1pt }$$
So the family of block-diagonal $2N\times2N$ matrices with $2\times2$ entry $\delta_j$, $1\leq j\leq4$ at the $i$:th place, $1\leq i\leq N$
$$\bigl(D^i_j\bigr)_{i\in\Lambda\cup\partial\Lambda}=\bigl(0\oplus\cdots\oplus\delta_j\oplus\cdots\oplus0
\bigr)\leqno(1.2)$$
gives an orthonormal basis (ONB) of 1-point functions $\widetilde{\cal O}_{\bf R}$.
\smallskip
\noindent {\it Example 2}:
$\widetilde{\cal O}_{\bf R}$ is the (real) algebra generated by Pauli matrices $(\widetilde D^i)_{i\in\Lambda\cup\partial\Lambda}$ with
diagonal block supported on site $i$ that takes values in
$\{\Id,i\sigma^x,i\sigma^y,i\sigma^z\}$. 
\smallskip
We shall restrict to the canonical algebra, whose generators enjoy the nice property of being real 
matrices.
Let also $o_{\bf R}\subset o$ be the algebra of $2\times 2$ matrices with real coefficients,
endowed with the scalar product $(A|B)=\Tra (B^*A)$, which is isometric with ${\bf R}^4$.
By extension, the basis $\delta=\{\delta_1,\delta_2,\delta_3,\delta_4\}$ of $o_{\bf R}$ will be called
an ``elementary basis'' of $\widetilde{\cal O}_{\bf R}$, since
$N$ copies of $\delta$, attached to each site $i$, give a basis $(D^i_j)_{i\in\Lambda\cup\partial\Lambda,1\leq j\leq4}$ of $\widetilde{\cal O}_{\bf R}$.
We say the same thing of any other ONB $b=\{b_1,b_2,b_3,b_4\}$ of $o_{\bf R}$, and of the corresponding basis
$(B^i_j)_{i\in\Lambda,1\leq j\leq4}$ of $\widetilde{\cal O}_{\bf R}$, where $B^i_j$ is defined as in (1.2), with $b_j$
instead of $\delta_j$.
Actually, the order of the elements of $b$ matters, so we prefer to think of $b$ as an ``array'', namely
$$b=\pmatrix{b_1\kern 1pt &b_2\kern 1pt \cr
b_3\kern 1pt & b_4\kern 1pt }\in{\cal M}_{4\times4}({\bf R})\leqno(1.3)$$
where $b_k$ is of the form
$$b_k=\pmatrix{b^{1k}\kern 1pt &b^{2k}\kern 1pt \cr
b^{3k}\kern 1pt & b^{4k}\kern 1pt }\in{\cal M}_{2\times2}({\bf R})$$
which we identify with the vector $b^k={}^t \bigl(b^{1k},b^{2k},b^{3k},b^{4k}\bigr)$.
We will not use the algebraic structure of $o_{\bf R}$.
With the notations above (partial traces), we see easily that~:
$$\tra _1(b)=\pmatrix{\tra b_1&\tra b_2\cr\tra b_3&\tra b_4\cr}\leqno(1.4)$$
which justifies the interpretation of $b$ as a matrix (operator).
So $\tra_1$ are the components, in some matrix representation, of the usual trace (``tracial state'') on $o_{\bf R}$.
For simplicity we set $T(b)=\tra _1(b)$ and call it the ``matrix of traces''.
An important r\^ole will be played with symmetric basis. 
\medskip
\noindent {\bf Definition 1.1}: 
{\it We call the ONB $b$ {\it symmetric} iff the corresponding matrix $b$ in (1.4) is Hermitian, i.e. $b_1=b_1^*$,
$b_4=b_4^*$, and $b_3=b_2^*$. We call it $\delta$-{\it symmetric} if moreover $b$ is real, and
$T(b)$ has a degenerate eigenvalue, that is, is a multiple of identity.}
\smallskip
Most of the basis are not symmetric, but occasionally we can make them symmetric,
by permuting or multiplying by $-1$ some elements. 
We can characterize $\delta$-symmetric basis: 
namely, if $b$ is $\delta$-symmetric,
then modulo such transformations, 
there exists discrete or one-parameter families $P_s\in O(2;{\bf R})$ such that 
$$b=b_s={}^tP_s\delta P_s\leqno(1.5)$$
(where the product is understood as if $b_j$'s were numbers). 

So far we have constructed ``one point functions'', i.e. a basis of $\widetilde{\cal O}_{\bf R}$. In the sequel we content
with Hamiltonians of type (0.1) which are of 
second order in the interactions; if we were to include the linear term
$\Sum_{i\in\Lambda}\sigma_i^z$ we would write it as $\Sum_{\langle i,j\rangle}1_i\otimes\sigma_j^z$.
Embed $\widetilde{\cal O}_{\bf R}$ into
$\widetilde{\cal O}_{\bf R}\otimes\widetilde{\cal O}_{\bf R}$ by the usual coproduct $\Delta$, and set
$\widetilde x=\Delta(x)={1\over2}(1\otimes x+x\otimes1)\in o_{\bf R}\otimes o_{\bf R}$, for $x\in o_{\bf R}$.
So we have ``lifted'' $\widetilde b=\Delta(b)$ as a family of $\widetilde{\cal O}_{\bf R}\otimes\widetilde{\cal O}_{\bf R}$
by $(\widetilde B^i_j)_{i\in\Lambda,1\leq j\leq4}$, with $\widetilde B^i_j=\Delta(B^i_j)$.
With the notations of (1.3) and (1.4) we have
$$\widetilde B^i=\pmatrix{\widetilde B^i_1\kern 1pt &\widetilde B^i_2\kern 1pt \cr
\widetilde B^i_3\kern 1pt & \widetilde B^i_4\kern 1pt }\in{\cal M}_{4N\times4N}({\bf R}), \quad
\tra _1(\widetilde B^i)=\pmatrix{\tra \widetilde B^i_1\kern 1pt &\tra \widetilde B^i_2\kern 1pt \cr
\tra \widetilde B^i_3\kern 1pt & \tra \widetilde B^i_4\kern 1pt }\in{\cal M}_{2\times2}({\bf R})
\leqno(1.6)$$
In the same way, we form $e^{-\beta H}\widetilde B^i_j$,
so we can map to each site $i\in\Lambda\cup\partial\Lambda$ a $2\times2$ matrix~:
$$\tra _1(e^{-\beta H}\widetilde B^i)=\tra (e^{-\beta H})\pmatrix{\omega_\beta(\widetilde B^i_1)\kern 1pt
&\omega_\beta(\widetilde B^i_2)\kern 1pt \cr
\omega_\beta(\widetilde B^i_3)\kern 1pt & \omega_\beta(\widetilde B^i_4)\kern 1pt }$$
\medskip
\noindent{\bf Definition 1.2}:
{\it We call {\it vorticity matrix} at site $i$, relative to the basis $b$, at inverse temperature $\beta$, the matrix~:
$$\Omega^i_\beta(b)={\tra _1(e^{-\beta H}\widetilde B^i)\over\tra (e^{-\beta H})}$$
The traceless matrix $$\widehat \Omega^i_\beta(b)=\Omega^i_\beta(b)-
\tra \bigl(\Omega^i_\beta(b)\bigr)\Id\leqno(1.7)$$
is called  the {\it reduced vorticity matrix} at site $i$.}
\medskip
\noindent {\it Example}: $\Lambda=\{1,2\}$ is a lattice with 2 sites, $\partial\Lambda=\emptyset$, one has
$\widehat \Omega^1_\beta(\delta)=\widehat \Omega^2_\beta(\delta)=0$. This is observed also numerically
for all $\Lambda$, with $\partial\Lambda=\emptyset$, although vortices could merge sponteanously in infinite volume.
\smallskip
If $b$ is a symmetric basis of $o_{\bf R}$, then $\Omega^i_\beta(b)$ and $\widehat\Omega^i_\beta(b)$
are hermitean since $H$ is self-adjoint (real symmetric if moreover $H$ has real coefficients), and 
$$\bigl(\widehat \Omega^i_\beta(b)\bigr)^2=\det\widehat \Omega^i_\beta(b)\Id\leqno(1.8)$$
Thus $\Omega^i_\beta(b)$ enjoys the nice property, to be diagonalizable with real (opposite) eigenvalues for all sites $i$,
and all inverse temperature $\beta$. Viewing these as a field of matrices over the lattice, we can figure out
the ``vorticity'' of the system, by simply looking at their principal directions.
We will see that it also gives a measure of vorticity, i.e. numbers (integers) that should be independent of the choice of ``elementary'' basis
$b$. Next we define vortices as the set of sites where the reduced vorticity matrix is singular.
\medskip
\noindent {\bf Definition 1.3}: {\it We say that $\xi\in\Lambda$ is a {\it vortex} at inverse temperature $\beta$
relative to the $\delta$-symmetric ONB $b$ iff
$\Omega^\xi_\beta(b)$ has a degenerate eigenvalue, i.e. $\widehat\Omega^\xi_\beta(b)=0$.
We call {\it regular} the other points.}
\smallskip
By construction, all sites are vortices when $\beta=0$.
\smallskip
Now we turn to consistency of Definitions 1.2 and 1.3 relatively to the choice of $b$ within $\delta$-symmetric basis.
That $b$ is a $\delta$-symmetric basis is a natural requirement for computing the degree of $\widehat\Omega^i(b)$, see Sect.2.
With $b$ written as in (1.3), and  $P\in O(2;{\bf R})$, we set with the notations of (1.5)
$a={}^tPbP$. 
The same holds after taking the co-product $\Delta$ of each term, i.e. $\widetilde a={}^tP\, \widetilde b P$.
This defines conjugacy classes, which pass to the partial traces (1.4), i.e. 
$T(a)={}^tPT(b)P$, and $T(\widetilde a)={}^tP\, T(\widetilde b)P$.
Moreover, if $X\in{\cal L}({\bf R}^2)$, we have 
$$(1\otimes X)b=\pmatrix{Xb_1\kern 1pt &Xb_2\kern 1pt \cr Xb_3\kern 1pt &Xb_4\kern 1pt}=(1\otimes X)Pa\, {}^tP\leqno(1.9)$$  
After lifting $\widetilde a$ and $\widetilde b$ to $\widetilde{\cal O}_{\bf R}\otimes\widetilde{\cal O}_{\bf R}$, 
(1.6) becomes
$$\widetilde A^i={}^tP\pmatrix{\widetilde B^i_1\kern 1pt &\widetilde B^i_2\kern 1pt \cr
\widetilde B^i_3\kern 1pt & \widetilde B^i_4\kern 1pt }P\in{\cal M}_{4N\times4N}({\bf R}), \quad
\tra _1(\widetilde A^i)={}^tP\pmatrix{\tra \widetilde B^i_1\kern 1pt &\tra \widetilde B^i_2\kern 1pt \cr
\tra \widetilde B^i_3\kern 1pt & \tra \widetilde B^i_4\kern 1pt }P\in{\cal M}_{2\times2}({\bf R})
\leqno(1.10)$$
and (1.9) also extends when taking $X\in{\cal L}({\bf R}^{4N})$ and replacing $a$ by $\widetilde A^i$, $b$ by $\widetilde B^i$. 
Let now $X=e^{-\beta H}$, we obtain that conjugacy classes pass to vorticity matrices, i.e. 
$$\Omega^i_\beta(a)={}^tP\Omega^i_\beta(b)P,\quad \widehat\Omega^i_\beta(a)={}^tP\widehat\Omega^i_\beta(b)P\leqno(1.11)$$
and (1.5) eventually gives~:
\medskip
\noindent {\bf Proposition 1.4}: {\it Definitions 1.2 and 1.3 are consistent, i.e. vorticity matrices relative to all
$\delta$-symmetric ONB $b$ are related by (1.11) for some $P_s\in O(2;{\bf R})$, and in particular
$\xi$ is a vortex relative to $\delta$ iff this is a vortex relatively to any $\delta$-symmetric $b$.}
\smallskip
Moreover we have the numerical evidence that, among all $\delta$-symmetric basis $b$, 
the canonical basis $\delta$ is most ``faithful'', in the sense that 
$\Omega^i_\beta(\delta)$ have on the boundary lattice $\partial\Lambda$ the same principal directions as the directions along which
Pauli matrices are compressed (associated with the eigenprojector $\Pi_j$). 
\medskip
\noindent {\it b) Heisenberg model}
\smallskip
The algebra goes essentially along the same lines, except for the fact that the basis $b$ cannot be real. 
Again, this relies on the observation that the spin representation of SU(2) is 2-D, so the ``one-point functions'' 
can be simply parametrized by $2\times2$ matrices.
Instead of $4\times4$ array
$b=\pmatrix{\delta_1&\delta_2\cr \delta_3&\delta_4\cr}$, we consider the $8\times8$ array
$e=\pmatrix{e_1&e_2\cr e_3& e_4}=\pmatrix{e'_1&e'_2+ie''_2\cr e'_3-ie''_3&e'_4\cr}$ with 
$$\eqalign{
&e'_1=\pmatrix{\delta_1&0\cr 0& 0\cr}, e'_2=\pmatrix{0&\delta_2\cr 0& 0\cr}, 
\quad e'_3=(e'_2)^* \cr
&e'_4=\pmatrix{0&0\cr 
0&\delta_4\kern 1pt \cr},  e''_2=\pmatrix{\delta_2&0\cr 0& 0\cr},  \quad e''_3=(e''_2)^*\cr
}\leqno(1.12)$$
this choice being non-unique. Moreover the $e_j$'s have the 
right dimension for quadratic interaction, so we don't need to take co-product as in the case of XY model.
Compression of Pauli matrices on $\partial\Lambda$ can be obtained by the orthogonal projector
$$\Pi(\theta,\varphi)=\pmatrix{\cos^2{\theta\over2}&e^{-i\varphi}\sin{\theta\over2}\cos{\theta\over2}\cr
e^{-i\varphi}\sin{\theta\over2}\cos{\theta\over2}&\sin^2{\theta\over2}\cr}\leqno(1.13)$$
where $(\theta,\varphi)$ parametrizes a point on Bloch sphere, accounting for the prescribed direction of ``vorticity'' on $\partial\Lambda$.

For both XY and Heisenberg model, we observe that (reduced) vorticity matrices belong to
$\su(2)=\{M\in{\cal M}_{2\times2}({\bf C}): M^*=M, \ \tra M=0\}$
which is the tangent Lie algebra of SU(2).
\medskip
\noindent {\bf 2. Topological degree and holonomy on SU(2)}. 
\smallskip
The natural idea is to consider vorticity matrices as a map $\Lambda\cup\partial\Lambda\to\su(2)$ and ``integrate'' it,
so to get some topologial invariant, such as the local degree. This assumes a thorough knowledge of discrete analysis
on the lattice, with values in su(2). For advanced results on Differential Calculus 
on lattices in the scalar case, see [Sm]. The non-commutative discrete case has still to be set up, so we 
pass here to an idealistic continuous limit, where vorticity matrices would be defined as a smooth field on ${\bf R}^2$
(away from vortices), 
valued in the Lie algebra su(2). Our purpose is to integrate such fields vanishing at some points,
and define a ``non-commutative degree''.

We consider Heisenberg model (Hermitean vorticity matrices), the XY model (real symmetric vorticity matrices)
will be treated as a particular case.
So let $M:D\subset{\bf R}^2\to\su(2), x\mapsto M(x)$ be a $C^1$ map, 
$M(x)^2=\lambda(x)\Id$, $\lambda(x)\geq0$, and consider 
$\rho\in\Lambda^1({\bf R}^2;\su(2))$ the 1-form
$$\rho(x)={1\over 2}(M^{-1}(x)dM(x)-dM(x)M^{-1}(x))\leqno(2.1)$$
(anti-symmetrized Maurer-Cartan form). We have
$M^{-1}(x)dM(x)+dM(x)M^{-1}(x))={d\lambda(x)\over\lambda(x)}$. 

Let $M=\pmatrix{a&b\cr c& -a\cr}$, we compute
$$d\rho=-\lambda^{-2}(a db\wedge dc+b dc\wedge da+ c da\wedge db)M\leqno(2.2)$$
(so $d\rho=0$ if $M$ is symmetric). On the other hand, computing 
the structure coefficients for the Lie algebra su(2), we find
$$d\rho+[\rho,\rho]=0\leqno(2.3)$$ 
Recall that if $G$ is a Lie group, 
and ${\cal G}$ its Lie algebra, $\omega$ the canonical 
Maurer-Cartan form on $G$, invariant by left translations, we 
define Darboux differential of the map $f\in C^1(D;G)$
by $\pi_f=f^*\omega$. The fundamental existence theorem (``Poincar\'e lemma''),
with a differential form
$\rho\in\Lambda^1(D;{\cal G})$ verifying (2.3), assigns (locally) a map $f\in C^1(D;G)$,  
whose Darboux differential is precisely equal to $\rho$. Moreover this map is unique
when prescribing its value at a point $x_0\in D$. Applying this result to (2.1), gives local primitives of $\rho$,
whenever $\lambda(x)\neq0$, called a ``logarithm'' of $M$, which belong to SU(2).  
\medskip
\noindent {\it Remark}: For these computations we can also use the isomorphism $\theta:({\bf R}^3,\wedge)\to(\su(2),{i\over2}[\cdot,\cdot])$, 
where $\wedge$ is the usual wedge-product on ${\bf R}^3$. 
\smallskip
For the XY model on ${\bf R}^2$ (``idealized'' lattice ${\bf Z}^2$) we can define the  square of the
``local degree'' at an isolated singularity (vortex) 
$\xi\in{\bf R}^2$, by
$$s_\xi^2=\det {1\over 2\pi}\int_{\gamma}\rho(x)\leqno(2.4)$$ 
where $\gamma$ encircles $\xi$, and similarly, 
when $M$ elliptic at infinity $|\lambda(x)|\geq C>0$, $|x|\geq r_0$
the square of the ``total degree''
$$s_\infty^2=\det {1\over 2\pi}\int_{|x|=r}\rho(x), \quad r>r_0\leqno(2.5)$$ 
Since the fundamental group of the universal covering of SU(2) is ${\bf Z}$, we can conclude that $s_\xi,s_\infty\in{\bf Z}$. 
\medskip
\noindent{\it Example}: For the symmetric matrices
$$M_a(x)=\pmatrix{a\cos n\theta &\sin n\theta \cr\sin n\theta & -a\cos n\theta }$$
we have $\deg _\infty(M)=n$. The 1-form $\rho$ associated wih $M_0$ is simply 
$\pmatrix{0&n \cr -n& 0\kern 1pt }d\theta$. 
\smallskip
This makes sense also for Heisenberg model on Riemann's sphere ${\bf S}^2$,
provided $\lambda(x)>0$ everywhere, but we should speak of  ``instantons'' rather than of ``vortices'',
see [BePo], [El-BRo]. 
\medskip
\noindent{\it Example}: For Hermitean matrices
$$M_a(x)=\pmatrix{a\cos\theta &e^{id\varphi}\sin\theta \cr e^{-id\varphi}\sin\theta & -a\cos\theta }$$
where $x=\pmatrix{\cos\theta &e^{i\varphi}\sin \theta \cr e^{-i\varphi}\sin\theta & -\cos\theta }\approx(\theta,\varphi)\in{\bf S}^2$,
we have $\deg _\infty(M)=d$. The condition $M(0,\varphi)=-M(\pi,\varphi)$ reproduces the condition of [BePo] that all (classical) spins
point upwards at infinity ($\theta=0$) while they point downwards at 0 ($\theta=0$), so that the equilibrium state at inverse
temperature $\beta$ is a spin wave, or ``instanton'', of degree $d$ (the number of coverings of the sphere).

Other topological defects, such as 
``lines of vortices'' occur in Heisenberg model
on ${\bf R}^3$ (``idealized'' lattice ${\bf Z}^3$). 

Among possible extensions, we mention: (1) the orbital compass model, which has reflection posivity, but no rotation invariance [BiChSt]; 
(2) Hubbard model with continuous symmetry (hopping and spin interaction) [KoTa], with 
an application to the dynamics of Cooper pairs in a supraconductor bulk, or the dynamics of electron/hole pairs in SNS junctions.
Vorticity can also defined for maps ${\bf R}^{(2n+1)n}\to\ssp (2n;{\bf R})$ (Hamiltonian matrices) and as well in general gauge sigma models [CiSa].
\medskip
\noindent{\bf 3. Numerical tests for the XY model}.
\smallskip
Recall we have completed the lattice $\Lambda$ with an environment $\partial\Lambda\subset{\bf Z}^2$
where Pauli matrices are compressed in directions $(\theta_j)_{j\in\partial\Lambda}$, i.e. we change $\sigma$ by
$\Pi_\theta\sigma\Pi_\theta$, where $\Pi_\theta$ as in (1.1). Thus
$$\sigma^x_i(\theta_i)=(\sin 2\theta_i)\Pi_{\theta_i}, \quad \sigma^y_i(\theta_i)=0$$
Hamiltonian (0.1) with nearest neighbor interaction has too large a kernel, to be suitable for
numerical simulations, even when modified by an external field. 
As in QFT we could try to remove the
``artificial'' part of $\ker H$ by reducing the Hilbert space ${\cal H}={\bf C}^{4N}$ to a ``physical space'', but $H$ is not positive in the form sense.
So a first attempt to lift the degeneracy of the spectrum 
of the Hamiltonian, and enhance the effects of the external field on vorticity, is to change (0.1) to the anisotropic XY model. So for  
$n,k>0$, we consider
$$\eqalign{
&H_{(n,k)}(\sigma|\partial\Lambda) =
-{1\over2(n+k)} \Sum_{\langle i,j\rangle ; i,j\in\Lambda}
({n}\, \sigma_i^x\otimes\sigma_j^x +
{k}\, \sigma_i^y\otimes\sigma_j^y)\cr
&-{1\over2(n+k)} \Sum_{\langle i,j\rangle ; (i,j)\in\Lambda\times\partial\Lambda}
{n}\, (\sigma_i^x\otimes\sigma_j^x(\theta_j)+
\sigma_j^x(\theta_j)\otimes\sigma_i^x)
-{1\over2(n+k)} \Sum_{\langle i,j\rangle ; i,j\in\partial\Lambda}
{n}\, \sigma_i^x(\theta_i)\otimes\sigma_j^x(\theta_j)\cr
}\leqno(3.2)$$
so $H_{(n,k)}(\sigma|\partial\Lambda)$ is self-adjoint and real.
For $k\neq1$, we call $H_{(1,k)}(\sigma|\partial\Lambda)$  the {\it anisotropic XY model}. Only when $\partial\Lambda=\emptyset$,
$H_{(1,k)}$ is unitarily equivalent to $H_{(k,1)}$.
We consider rectangular lattices of minimal sizes to exclude important volume effects, with sufficiently large $\partial\Lambda$
to constrain the ``quantum system'' within $\Lambda$. We choose $\theta_j=d\omega_j+\phi$ where $\omega_j$ is the polar angle
representing the vector $j\in\partial\Lambda$. 

We study Gibbs state at inverse temperature $\beta$, with significant results provided $\beta$ ranges in some interval, for which
however, there is no evidence (even in an approximate sense) of a second order phase transition.
Computing ${1\over 2\pi}\int_\gamma\rho$
as a discrete integral along a contour $\gamma\in\Lambda$,
not too far from $\partial\Lambda$ (in practice, 2 or 3 layers),
it turns out that the computed degree is close
to this we would obtain in (2.6). 
The main flaw affecting the computations is due to the fact that eigenvalues of $\widehat \Omega^i_\beta(b)$
are decaying exponentially when approaching the center of $\Lambda$.
This we partially compensate by considering the anisotropic model.
To make our vorticity patterns more demonstrative we consider the case of high anisotropy with $k=10$. 
In Table~1 
below we give the results for degree, computed along a cycle $\gamma\subset\Lambda$ 
consisting of the rectangle of the first or the second neighbors to the boundary for different values of the anisotropy parameter $n=1,k=2; 10$.
Inverse temperature is $\beta=1$.
\medskip
{\it Table 1.} Table of calculated degree
for different values of anisotropy factor $k$
for the first and the second neighbors to the boundary.
Here the case of $|\Lambda\cup\partial \Lambda| = 23\times 33$ with 2 boundary layers
is considered, and $\beta=1$.
$$\matrix{Given \,\, & k=2\,\, & k=10 \,\, & k=2\,\, & k=10\cr
degree &\, 1^{st} \, neighbours &\, 1^{st} \, neighbors & \, 2^{nd} \, neighbours 
& \, 2^{nd} \, neighbours   \cr
1&1.05&1.09&0.89&1.05\cr
2&1.98&2.03&1.70&1.78\cr
3&2.76&2.75&2.01&2.50 }
$$
\medskip
Let us finally discuss the antiferromagnetic model. It is known that on ${\bf Z}^2$,
the unitary transformation $U$ consisting in flipping the spins at sites $i$
with $i$ odd (i.e. indices $i=(i_1,i_2)$ such that $|i|=|i_1|+|i_2|$ is odd)
intertwines the ferro with the antiferromagnetic models. More precisely, $-H=U^*HU$. The reason is that
${\bf Z}^2_e$ and ${\bf Z}^2_o$
(the even and odd lattices) are swapped into each other by symmetries on the lines
$x=n+1/2$ or $y=m+1/2$ (called the ``chessboard symmetry'').
There follows that $\tra \exp[\beta H]A=\tra \exp[-\beta H]UAU^*$, and if $A=\widetilde D^i$ (the canonical basis),
we can check $UAU^*=A$ so the matrices of vorticity (for the Hamiltonian with free boundary conditions) are the same.
This equivalence holds also in the case of the torus, but not on $\Lambda\subset{\bf Z}^2$ with an odd number of sites.
Of course, when $\partial\Lambda\neq\emptyset$, $H$ and $-H$ are not so simply related; nevertheless, we may observe (numerically) that the relation
$\Omega^i_\beta(\delta)=\Omega^i_{-\beta}(\delta)$ holds with a very good accuracy.
\medskip
\noindent {\bf References}:
\smallskip
\noindent [AsPi] W.Aschbacher, C.A.Pillet. Non-equilibrium steady states of the XY chain. J. Stat. Phys. 112, p.1153-1175, 2003.

\noindent [BePo] A.Belavin, A.Polyakov. Metastable tates of 2-d isotropic ferromagnets. JETP Lett., Vol 22, No.10, p.245-247, 1975.

\noindent [BiChSt] M.Biskup, L.Chayes, S.Starr. Quantum spin systems at finite temperature. Commun. Math. Phys.
Vol 269, 3, p.611-657, 2007. 

\noindent [BrFoLa] J.Bricmont, J.Fontaine, J.Landau. On the uniqueness of the equilibrium state for the plane rotator.
Commun. Math. Phys. 56, p.281-286, 1977.

\noindent [CiSa] K.Cieliebak, D.Salamon. The symplectic vortex equations and invariants of Hamiltonian group actions.
J. Symplectic Geom. 1(3), 543–645, 2002.

\noindent [El-BRo] H.El-Bouanani, M.Rouleux. Vortices and magnetization in Kac's model. 
J. Stat. Physics. Vol.28, No.3, p.741-770, 2007. 

\noindent [FrLi] J.Fr\"ohlich, E.Lieb. Phase transitions in anisotropic lattice spin systems. Commun. Math. Phys. 60, p.233-267, 1978.

\noindent [FrSp] J.Fr\"ohlich, T.Spencer. The Kosterlitz-Thouless phase transition in 2-D Abelian spin systems and the Coulomb gas.
Commun. Math. Phys. 81, p.527-602, 1981. 

\noindent [KoTa] T.Koma, H.Tasaki. Decay of superconducting and magnetic correlations in 1-D and 2-D Hubbard model, 
Phys. Rev. Letters 68(21), p.3248-3251, 1992.

\noindent [Ma] P.Malliavin. G\'eom\'etrie Diff\'erentielle Intrins\`eque, Hermann, Paris, 1972.

\noindent [MeMiPf] A.Messager, S.Miracle, C.Pfister. Correlation inequalities and uniqueness of equilibrium state for the planar
rotator ferromagnetic model. Commun. Math. Phys. 58, p.19-29, 1978.

\noindent [Sm] S.Smirnov. Discrete complex analysis and probability. 
http://www.unige.ch/~smirnov/slides/sli des-hyderabad.pdf
\bye